Progress in the definition of a reference human mitochondrial proteome*

Pierre Lescuyer[1], Jean-Marc Strub[2], Sylvie Luche[1], Hélène Diemer[2], Pascal Martinez[1], Alain Van Dorsselaer[2], Joël Lunardi[1] and Thierry Rabilloud[1,3]

1) CEA- Laboratoire de BioEnergétique Cellulaire et Pathologique, UPRES 2019/UJF
DBMS/BECP
CEA-Grenoble, 17 rue des martyrs
F-38054 GRENOBLE CEDEX 9 FRANCE

2) Laboratoire de Spectrométrie de Masse Bio-Organique, UMR CNRS 7509, ECPM, 25 rue Becquerel, 67087 STRASBOURG CEDEX 2

3) to whom correspondence should be addressed

*: This work can be found on the Internet at the following address:

http://www-dsv.cea.fr/Thema/MitoPick/Default.html

(Running title): Human mitochondrial proteome

Correspondence :

Thierry Rabilloud, DRDC/BECP
CEA-Grenoble, 17 rue des martyrs,
F-38054 GRENOBLE CEDEX 9
Tel (33)-4-76-88-32-12
Fax (33)-4-76-88-51-87
e-mail: Thierry.Rabilloud@ cea.fr

Summary

Owing to the complexity of higher eukaryotic cells, a complete proteome is likely to be very difficult to achieve. However, advantage can be taken of the cell compartmentalization to build organelle proteomes, which can moreover be viewed as specialized tools to study specifically the biology and "physiology" of the target organelle. Within this frame, we report here the construction of the human mitochondrial proteome, using placenta as the source tissue. Protein identification was carried out mainly by peptide mass fingerprinting. The optimization steps in 2D electrophoresis needed for proteome research are discussed. However, the relative paucity of data concerning mitochondrial proteins is still the major limiting factor in building the corresponding proteome, which should be a useful tool for researchers working on human mitochondria and their deficiencies.

Abbreviations

BisTris: N-N bis (hydroxyethyl) N-N-N tris (hydroxymethyl) aminomethane; CHAPS: 3[ (3-cholamidopropyl) dimethylammonio] propane sulfonate; CID: Collision-Induced Degradation; ECL: enhanced chemiluminescence; IEF: Isoelectric focusing; IPG: IEF in immobilized pH gradients; IPG-DALT: 2D electrophoresis with IEF in immobilized pH gradient in the first dimension and SDS PAGE in the second dimension; MALDI: Matrix Assisted Laser Desorption and Ionization; MS: mass spectrometry; PAGE: polyacrylamide gel electrophoresis; PBS: phosphate buffered saline; PSD: post-source decay; TOF: time of flight

1. Introduction

The classical approach in proteomics couples 2D electrophoresis with post-gel identification by mass spectrometry. While this technique has proved quite efficient, as shown by the numerous databases available to date (see the Expasy server at www.expasy.ch/ch2d/2d-index.html for a partial index of such databases), its limitations are now well described [1]. Among those limitations, three are felt with increasing acuity: (i) the difficulty in the analysis of basic and large proteins [2], (ii) the under-representation of membrane proteins [3] and (iii) the poor analysis of low-abundance proteins. The latter limitation is due both to the high dynamic range of protein expression in living cells and to the relative lack in resolution and capacity of 2D gels. While the latest techniques used for 2D gels seem to be able to cope with the proteome complexity of prokaryotes [4, 5], they still do not match with the complexity encountered in multicellular eukaryotes. One way to circumvent this difficulty is to cut the complex sample represented by the eukaryotic cell into several less complex samples. This can be done on chemical criteria (e.g. narrow pH gradients, as in [2] and [5]) but also on biochemical criteria, using the compartimentalization of eukaryotic cells. This latter approach adds valuable localization and functional information and is therefore quite attractive. One of the first example of such subcellular proteomes has been made on mitochondria [6], but other, more recent examples have been shown on lysosomes [7], golgi [8] or chloroplasts [9].

While this work on organelle proteomes has demonstrated its interest both on the functional point of view and its ability to deal with lower abundance proteins, it is becoming apparent that the proteome coverage is still far from perfect. In addition to the technical limitations for basic, large or hydrophobic proteins, the dynamic range is still not sufficient and further fractionation of the organelle content has been proposed [10]. However, these fractionation techniques require massive amounts of mitochondria, which are not easily available either from patient biopsies or even from model cell culture systems [11], [12]. This is why we chose to use only more straightforward proteomic approaches.

However, proteomic techniques are still evolving, and some of the above-mentioned limitations have been addressed by dedicated gradients for basic proteins [13], improved solubilization cocktails, detection methods or mass spectrometry methods or apparatus. Keeping to the mitochondrial example, we have therefore decided to investigate how these technical improvements can help in improving the coverage of organelle proteomes.

2. Materials and methods

2.1. Isolation of placental mitochondria

The method of preparation of mitochondria from human placenta was modified from Ausenda and Chomyn [14]. Human placenta were obtained from the local hospital (Grenoble) within 1 h of delivery and transported in ice to the laboratory. All manipulations were performed at 4°C unless otherwise stated. The tissue mostly freed of membrane and connective tissue was cut in small pieces and washed extensively by 2 L of 10 mM phosphate-buffered saline pH 7.4 (138mM NaCl, 2,7 mM KCl). The rinsed and drained tissue was then processed for 1 min with 500 ml homogeneization buffer (0.25 M sucrose, 0.15 M KCl, 10 mM Tris-Cl, pH 7.5, 1 mM EDTA, 0.5% fatty acid-free BSA) in a Waring blender at low speed. Homogenate was centrifuged 10 min at 800g. The collected supernatant was filtered through layers of cheesecloth and centrifuged 15 min at 12000g. The resulting pellet was resuspended in 100 ml of 0.225 M sucrose, 0.075 M mannitol, 10 mM Tris-Cl, pH 7.5, 1 mM EDTA) and centrifuged 10 min at 1000g. The mitochondria were then pelleted by centrifuging the 1000g supernatant at 12000g for 15 min, resuspended in the same buffer and washed twice. Mitochondria were stored in liquid nitrogen.

2.2. Sample preparation for 2D gels.

The mitochondrial suspension was diluted directly fivefold in a concentrated lysis solution containing 9M urea, 2.5M thiourea, 5% (w/v) CHAPS, 12.5 mM DTT and 0.5% (w/v) carrier ampholytes (Pharmalytes 3-10) [15], or in some instances in a solution containing other detergents [16]. After 1 hour at room temperature, the solution was directly used for in gel sample-rehydration.

2.3. Immobilized pH gradient focusing

2.3.1. Gel casting

The Immobiline concentrations used to generate the pH gradients (mean buffering power = 3 mequiv.$l^{-1}$.$pH^{-1}$) were calculated according to published recipes [17]. The pH 5.5 to 12 gradient is interpolated from the previously described 4-12 gradient [13]. Linear pH gradients with plateaus were used in all cases [18]. Gels were cast at 4.5%T, and polymerized at 50°C.

2.3.2. IPG strip rehydration, running and equilibration

4mm wide IPG strips were cut from the dry plate with the use of a paper cutter. The protein solution was added to the denaturing solution (4% (w/v) CHAPS, 0.5% (v/v) Triton X100, 0.4% (w/v) carrier ampholytes (Pharmalytes 3-10), 10mM DTT, 2M thiourea and 7M urea) to a final volume of 400 $\mu l$ (sample loading with cup) to 600 $\mu l$ (sample loading by rehydration). In some cases, a solution containing other detergents than CHAPS was used.

The strips were run on a Dry-strip kit according to the manufacturer's instructions with the previously described modifications [18]. The entire set-up was covered with low viscosity silicon or paraffin oil, and the temperature set at 22°C with a circulating water bath.

Migration was carried out for a total of 50 to 60 kVh [15].

2.4. Equilibration, SDS dimension and staining

After the IEF run, the oil was poured off, and the strips equilibrated while in place in the running setup, as previously described [19]. The strips were then sealed on the top of the 1.5mm thick second dimension gel (Bio-Rad vertical system) with the help of 1% low-melting agarose in 0.2% SDS, 0.15M BisTris-0.1M HCl buffer supplemented with bromophenol blue as a tracking dye.

SDS-PAGE was carried out at constant power (15W per gel) with cooling at 10°C, until the tracking dye reached the bottom of the gel. Gels were stained with silver [20]. Basic gels were stained with an ammoniacal silver staining with either a first fixation in 4% formaldehyde-25% ethanol [21] or fixation in 0.05% naphthalene disulfonate in 5% acetic acid and 30% ethanol [6]. Preparative gels for spot excision for mass spectrometry were generally stained with a home-made fluorescent ruthenium complex [22].

2.5. In-gel protein digestion and MALDI-MS

Stained proteins spots or bands were excised (on a UV table for fluorescent detection). Silver-stained gel pieces were destained with ferricyanide-thiosulfate [23]. Gel pieces were then shrunk in 1 ml of 50% ethanol for 2 hours, and stored at -20°C.

Each gel slice was cut into small pieces with a scalpel, washed with 100 $\mu$l of 25 mM NH4HCO3, agitated for 8 min with a Vortex mixer. After settling of the gel pieces, the supernatant was removed. Gel pieces were dehydrated with 100 $\mu$l of acetonitrile for 8 min. This operation was repeated twice. Gel pieces were completely dried with a Speed Vac (15 min.) before reduction-alkylation. Gel pieces were covered with 100 $\mu$l of 10 mM DTT in 25 mM NH4HCO3 and the reaction was left to proceed at 57°C for 1 hour. The supernatant was removed, 100 $\mu$l of 55 mM iodoacetamide in 25 mM NH4HCO3 were added and reaction was left in the dark at room temperature for 1 hour. The supernatant was removed and the washing procedure with 100 $\mu$l of NH4HCO3 and acetonitrile was repeated three times. Gel pieces were completely dried with a Speed Vac before tryptic digestion. The dried gel volume was evaluated and three volumes of trypsin (12.5 ng/$\mu$l) in 25 mM NH4HCO3 (freshly diluted) were added. The digestion was performed at 35°C overnight. The gel pieces were centrifuged and 5 $\mu$l of 25% H2O/70% Acetonitrile/5% HCOOH were added. The mixture was sonicated for 5 min. and centrifuged. The supernatant was recovered and the operation was repeated once. The supernatant volume was reduced under nitrogen flow to 4 $\mu$l, 1 $\mu$l of H2O/5% HCOOH were added and 0.5 $\mu$l of the mix were used for MALDI-TOF analysis.

Mass measurement were carried out on a Bruker BIFLEX™ MALDI-TOF equipped with the SCOUT™ High Resolution Optics with X-Y multisample probe and gridless reflector. This instrument was used at a maximum accelerating potential of 20 kV and was operated in reflector mode. A saturated solution of α–cyano-4-hydroxycinnamic acid in acetone was used as a matrix. A first layer of fine matrix crystals was obtained by spreading and fast evaporation of 0.5 $\mu$l of matrix solution. On this fine layer of crystals, a droplet of 0.5 $\mu$l of aqueous HCOOH (5%) solution was deposited. Afterwards, 0.5 $\mu$l of sample solution was added and a second 0.2 $\mu$l droplet of saturated matrix solution (in 50% H2O/50% acetonitrile) was added. The preparation was dried under vacuum. The sample was washed one to three times by applying 1µl of aqueous HCOOH (5 %) solution on the target and then flushed after a few seconds. Internal calibration is performed with Angiotensin 1046.542 Da, Substance P 1347.736 Da, Bombesin 1620.807 Da, and ACTH 2465.199 Da.

Monoisotopic peptide masses were assigned and used for databases searches. The MASCOT program was used to database searching [24]. All proteins present in NCBInr were taken into account without any species, pI and Mr restrictions. The peptide mass error was limited to 100 ppm, 1 missed cleavage might be accepted and no AA substitutions were allowed. The statistical tools provided by MASCOT as well as the species of the candidate proteins returned by MASCOT were used for result interpretation and protein identification. For example, MOWSE score greater than 75 only were considered as significant, and at least 3 matching peptides were necessary for a positive identification. This rather low number was dictated by our need to identify low molecular weight proteins.

3. Results

3.1. 2D electrophoresis

Immobilized pH gradients were selected for their intrinsic reproducibility and loading capacity. In addition, they should allow analysis of basic proteins [13]. To this purpose, we decided to complement our standard 4-8 map with basic proteins. To this purpose, we used either wide pH gradients [25] or more dedicated gradients. As the very basic pH gradients (8-12, 9-12) seem rather difficult to use [26], we decided to cast a pH 5.5 to 12 pH gradient, starting from the published 4-12 pH gradient [13]. For this gradient, as well as for the other basic ones, anodic cup loading proved much more efficient than in gel sample rehydration [25, 27]. Oppositely, the focusing time was found not to be critical, so that we used our standard 24 hours migration, resulting in ca. 60 kVh focusing, for 4-8, 3-12 and 5.5-12 gradients. The optimal resolution was reached by using a 4-8 and a 5.5-12 pH gradient. As shown on figure 1, the pH 3-12 gradient led to important compression, and therefore low resolution, in the acidic part. The basic part of this gradient showed quite nice focusing and spot definition, but the actual pIs of the proteins displayed on this gradient (e.g. MDHM or POR1) did not extend above pH 9, while it should extend at least to pH 11. Therefore, many basic mitochondrial proteins are missing on those wide pH gradients (compare figure 1 and figure 3). Oppositely, the 5.5-12 gradient showed greater streaking, but improved resolution and protein display above pH 9.

3.2. Reference map analysis and protein identifications

The Melanie software detected ca. 1500 spots on the silver-stained acidic reference map gel, i.e. in a pI 4 to 8 and a Mr 10-200 kDa range. 1250 spots were detected on the basic reference gel (pH 5.5-12 range). The dynamic range varied from 30 to 7500 arbitrary detection units, for a total integrated density of 350000. This means that proteins present at 0.01 % of the initial input can be visualized (see figure 3). However, the less abundant proteins identified to date by general methods (i.e. neither comigration nor blotting) represent 0.03 % of the original sample, while the majority of medium spots identified are in the 0.1-0.5% range.

The reference 2D map of human placenta mitochondria and the current state of the corresponding proteome are shown on figures 2 and 3 and Tables 1 and 2. Spot identification was easy by MALDI-TOF above 15-20 kda. Below this mass, a low number of peptides was often encountered, which precluded secure protein identification. As an example, COX Va, Vb and VIa could be identified by MALDI TOF, while ATP synthase δ and thioredoxin required MS/MS for safe identification [6, 11]. However, it must be mentioned that basic proteins, richer in lysine and arginine than their acidic counterparts, were much easier to identify with MALDI, even with low molecular weight proteins. An example is shown in Figure 4 with the MALDI spectrum of the NUML/MLRQ subunit of complex 1. For this small, basic protein, 4 peptides were obtained, which is a small number, but the sequence coverage is over 70% (see table 1), which affords secure protein identification. Conversely, mitochondrial thioredoxin, which is a larger but more acidic protein, only gave 2 tryptic peptides which did not provide safe protein identification.

4. Discussion

4.1. Protein detection optimization.

As soon as peptides need to be extracted from the 2D gel for the purposes of spot identification, the detection method needs to be optimized. For example, even with glutaraldehyde-free protocols, which are mass spectrometry compatible, silver staining results in peptide losses and therefore decreased identification efficiency [28]. For this reason, many groups use colloidal Coomassie blue staining at least for micropreparative work. In some instances (e.g. [4]) this method is the only detection method used. However, mitochondrial proteins are much less easy to obtain than bacterial proteins, so that such an approach cannot be used in our case. In addition, the detection threshold of colloidal Coomassie blue is rather high [29]. This ensures that enough protein is present for mass spectrometry, but this conversely requires very high loads to analyze minor proteins. Moreover, the improvements in identification methods allow to identify proteins at levels which are below the colloidal Coomassie blue detection threshold. For this reason, we used zinc-imidazole staining for micropreparative work and silver staining for the analytical work in our previous study [6]. However, zinc-staining provides very low contrast, resulting in major difficulties for spot assignment and excision. We therefore decided to use a more user-friendly method for our micropreparative work. Fluorescent detection, using either commercial probes [30, 31], or home-made ones [22], is known to provide sensitive detection and low interference with mass spectrometry. We decided to use a home-made ruthenium complex for our micropreparative work, as it combines high sensitivity, low cost, minimal interference with mass spectrometry, and easy excitation with a UV table for spot excision [22].

Analytical detection was carried out with silver staining, using a silver nitrate method [20] for acidic and neutral proteins (i.e. 4-8 and 3-12 gradients) and ammoniacal silver methods [6], for alkaline proteins (i.e. 5.5-12 gradients). It must be emphasized here that some basic proteins stain nicely with ammoniacal silver staining with formaldehyde fixation but are hardly detectable with any other method [21]. The ADP-ATP carrier, a major inner membrane protein, is a typical example of such proteins.

4.2. Scope of analysis

(i) The classical problem, when dealing with an organelle proteome, is the one of the contamination by other cell components. This cannot be ruled out, whatever the purification method of the organelle may be. A typical example in the case of placenta mitochondria is shown by the choriomammotropin (PLL). This protein is a major secreted protein, and is found at high levels in mitochondrial preparations (see figure 2). This is probably due to the fact that the corresponding secretion granules have a rather large size and a density close to that of mitochondria. The corresponding spot is of course absent when mitochondria from other origins (e.g. cultured cells) are analyzed. However, as they represent only 10% of the protein content of the cells, the overall protein yield is low. This makes the use of cultured cells critical to obtain the protein amounts still needed for reference proteome work, while they can be invaluable models in other aspects [11, 12]. Thus, for reference purposes, it is much better to start from tissues, and this explains why bovine heart has been used for years in mitochondrion research. Another obvious contaminant is the BSA used during the initial isolation stages of the mitochondria, which is found in the 2D maps, even if the mitochondria are washed several times in BSA-free buffer. The other contaminants observed in our mitochondrial proteome may be linked to the mitochondrion-cytosol interface. This is obviously the case for cytoskeletal proteins (e.g. actin and Rho protein) which probably act as positioners for the mitochondria within the cell and are therefore linked to the surface of the mitochondria. However, typical ER proteins are also encountered, such as a diaphorase isoform, disulfide isomerases and some chaperones (grp78 and 94). This shows a cross contamination of our mitochondrial preparations with ER. Such cross contaminations greatly vary from one source to another, and are for example much less pronounced with heart or liver tissue than with placenta. However, when dealing with human tissues, ethical issues apply. This is why we chose to use placenta as a starting tissue, despite the fact that placental mitochondria exhibit greater contamination than heart mitochondria, as shown here and in previous work [6]. We tried to remove these non-mitochondrial contaminants by using density or viscosity gradients, but the gain in purity was marginal, while the yield decreased severely.

(ii) Several protein spots gave good spectra but no hits in the databases. This is most probably linked to the rather low number of mitochondrial proteins described in the databases. A search in the Swiss-Prot database yielded only 311 entries of human mitochondrial proteins. This database coverage problem is also examplified by the fact that peptide mass fingerprinting sometimes gives a hit for recently cloned genes (e.g. Q9HBL7) for which very few functional data are available.

(iii) This problem is however attenuated by the fact that mitochondrial proteins seem to be very highly conserved from one species to another. This is shown for example from the identification of ATPase subunits or heat shock proteins . In such cases, the first hit is the correct protein, but the homologous proteins from other species also give very high scores. This allows cross-species identification by peptide mass fingerprinting with rather good confidence. This has been already described for mouse ATP synthase beta [32], and is probably also the case in this study with the first description of a human sideroflexin protein.

(iv) basic and small proteins

One of the problems associated with the analysis of mitochondrial proteins is that they are general more basic than the standard cytosolic proteins. This was a major problem up to the very last years. However, the appearance of robust basic immobilized pH gradients [13] dramatically changed the situation and allowed to display proteins with a pI up to 10 without major problems. The situation is still difficult above pH 10, as shown by the absence of mitochondrial ribosomal proteins on our maps. However, in addition of being very basic, mitochondrial ribosomal proteins are also very weakly abundant [33], which further hampers their analysis in total mitochondrial extracts.

Mitochondria are also very rich in low-molecular weight (<10 kDa) proteins. This is the case for example in the respiratory complexes. NADH ubiquinone oxidoreductase (respiratory complex I) is composed of more than 40 subunits, and 7 are below 10 kDa. Quite often, these small proteins are also basic and completely escaped from the analysis in our previous work. Despite the fact that they are quite difficult to detect, we could analyze some of them, as examplified by the detection of the MLRQ, B17 and B18 subunits of complex I. The same holds true for ATP synthase, where only 2 subunits out of 13 were detected in our previous work [6], while 6 subunits have been now identified on our 2D maps.

(v) membrane proteins

Membrane proteins are known to be a major issue in 2D PAGE-based proteomics [3]. This is examplified by the fact that even not a single membrane proteins was described in our previous mitochondrial proteome. This is fortunately no longer the case, thanks to significant advances in the solubilization of hydrophobic proteins for 2D PAGE [16]. Thus, true membrane proteins can be seen in the current proteome, even with basic pIs. It must be mentioned, however, that mitochondrial membrane proteins belong to three different classes. The first class is the one of beta barrels proteins [3], which includes the porins (mitochondrial or bacterial). These proteins are easily seen on 2D maps, even with standard conditions [34], and are therefore easily detected on our maps. The second class is represented by the respiratory complexes. In these complex protein assemblies, different subunits build the membrane-embedded anchor and the mitochondrial matrix-protruding extension. A good example of this situation is ATP synthase, which is made of a membrane-embedded Fo domain and a protruding F1 extension. Quite often, the membrane-embedded subunits are small and very hydrophobic, a poor situation for 2D PAGE. However, we could detect some of the Fo subunits (ATP synthase e and g chains). Many of these proteins are still missing to date (e.g. other Fo subunits and the ND1 to ND6 subunits of complex I), but further progress is still expected. The last class of mitochondrial membrane proteins is made of more classical, transmembrane helices-containing proteins, here again quite often basic. Rather surprisingly, this class is the one which performs less in our system. For example, we could not detect and analyze the glutamate-malate exchanger, the phosphate carrier etc..., which represent a subclass of basic (pI ca. 10) medium-sized (30 kDa) proteins. We could however analyze two members of this class. One is the ATP-ADP carrier, a very abundant member of this class and one other is very likely to be sideroflexin, i.e. a putative iron or iron-related metabolite transporter [35]. This first successes suggest that we may be close to be able to handle these proteins, likely by further improvement through the use of optimized detergent-chaotropes combinations. As previously described [3], the ratio between the size of the protein and the number of transmembrane segments seems to be the key parameter. As a positive example, mitofilin, which is a 100 kDa protein with a single predicted transmembrane segment [36], is easily analyzed in our system (Figure 2). Although it is almost sure that some membrane proteins will escape 2D PAGE-based proteomics, the improvements achieved in the last years (from 0 to 10 mitochondrial membrane proteins) make us confident that the situation will improve further. It must be mentioned, however, that a further difficulty in the analysis of membrane proteins is represented by their limited abundance, even in mitochondria. In the analysis of membrane proteins by 2D electrophoresis followed by mass spectrometry, it must be noted that peptides corresponding to segments of the transmembrane helices can be seen in the mass spectra.

4.3. Concluding remarks

One of the major problems when dealing with organelles concerns the delimitation of what belongs to the organelle and what does not. However, most of the so called "contaminants" which we encountered during our study may well be at least loosely bound to the outer surface of the mitochondria. In addition, it must be noted that proteins devoid of transit peptides may be present both outside and inside the mitochondria, as shown by the example of glutathione reductase [37].

The other big challenge remaining is the one of membrane proteins, as is the case for total extracts. Other, non 2D-based approaches are probably more suitable to point out membrane proteins present in the sample (e.g. the MUDPIT approach [38] and the ICAT approach [39]) . These peptide-based approaches also do not differentiate between the complete protein and its fragments, which can be quite abundant, as shown by the number of ATP synthase fragments present on our maps. To alleviate this problem, SDS PAGE-based approaches [8] appear more suitable. However, all these approaches do not provide easy access to the modified forms of the proteins (see the mitofilin profile) and to the intra-sample relative abundance of the proteins, which 2D PAGE does.

Acknowledgments:
TR wishes to thank CNRS for personal support.

Figure legends

Figure 1: 2D electrophoresis of mitochondrial proteins on a wide pH range.
0.2 mg mitochondrial proteins are loaded on 4mm wide IPG strips (pH interval 3-12),
in a solution containing 7M urea, 2M thiourea, 4% CHAPS, 0.4% carrier ampholytes (3-10 range) and 20 mM DTT. Second dimension 10%T SDS gel. Detection by silver tetrathionate [20].

Figure2 : Acidic and neutral mitochondrial proteins.
The mitochondrial proteins are separated on a 4-8 linear pH gradient. Sample is loaded by in gel rehydration. Sample solution as in figure 1. (A) 10% T SDS gel in the second dimension. (B) 12%T SDS gel in the second dimension. The proteins are identified via their name in the SwissProt database or their TrEmbl number (see also table 1). Red arrows: established mitochondrial proteins. Green arrows: fragments of established mitochondrial proteins. Blue arrows: poorly characterized proteins (TrEmbl numbers only). Black arrows: non mitochondrial proteins.

Figure3 : Basic mitochondrial proteins.
The mitochondrial proteins are separated on a 6-12 non-linear pH gradient [13]. Sample is loaded by cup loading on the anodic side. Sample solution: 7M urea, 2M thiourea, 2% deoxyCHAPS, 0.4% carrier ampholytes (3-10 range) 20 mM DTT and 5mM tris cyanoethyl phosphine (Molecular Probes). (A) 10% T SDS gel in the second dimension. (B) 12%T SDS gel in the second dimension. The proteins are identified via their name in the SwissProt database or their TrEmbl number (see also table 1). Red arrows: established mitochondrial proteins. Green arrows: fragments of established mitochondrial proteins. Blue arrows: poorly characterized proteins (TrEmbl numbers only). Black arrows: non mitochondrial proteins. Orange arrow: protein present at a 0.01% level.

Figure 4: MALDI mass spectrum of trypsin-digested NUML/MLRQ subunit of complex 1 and of the ATP-ADP exchanger.
The MALDI mass spectra of two mitochondrial proteins is shown. A: NUML/MLRQ, B: ATP-ADP exchanger. The peptides used for identification are marked with a star. NUML/MLRQ is a small protein (Mw <10 kDa) while the ATP-ADP exchanger is a typical, multi transmembrane helix-containing inner membrane protein. The peptides shown by an arrow correspond to segments of the transmembrane helices: 1219.63 (189-199), 1446.74 (81-92), 1755.99 (281-296), 1927.09 (172-188)

| Code | Name | Accession number | mass | pI | Sequence coverage (%) |
|---|---|---|---|---|---|
| D3HI | 3 hydroxybutyrate dehydrogenase | P31937 | 31 537 | 5.54 | 37 |
| AATM | Aspartate aminotransferase | P00505 | 44 695 | 8.98 | 31 |
| ACDS | acyl CoA dehydrogenase short-chain | P16219 | 41 721 | 6.15 | 29 |
| ACDM | Acyl-CoA dehydrogenase, medium chain specific | P11310 | 43 643 | 7.02 | 32 |
| ACDV | Acyl-CoA dehydrogenase, very-long-chain specific | P49748 | 66 175 | 7.74 | 32 |
| ACON | Human aconitase | Q99798 | 82 426 | 6.85 | 11 |
| ADT2 | ADP, ATP carrier protein, ANT2 | P05141 | 32 895 | 9.76 | 36 |
| ATPA | ATP synthase alpha chain | P25705 | 55 209 | 8.28 | 45 |
| ATPB | ATP synthase, beta chain | P06576 | 51 769 | 5.00 | 20 |
| ATPD | ATP synthase, delta chain | P30049 | 15 020 | 4.53 | ND |
| ATPE | ATP synthase E chain | AAH03679 (NCBI) | 7 928 | 9.34 | 72 |
| ATPN | ATP synthase G chain | O75964 | 11 386 | 9.60 | 33 |
| ATPO | ATP synthase oligomycin sensitivity conferral protein | P48047 | 20 875 | 9.81 | 47 |
| C11A | Cytochrome P450 11A1 | P05108 | 56 117 | 7.94 | 38 |
| CH60 | 60 kDa heat shock protein | P10809 | 57 963 | 5.24 | ND |
| COXA | Cytochrome c oxidase polypeptide Va | P20674 | 12 513 | 4.88 | ND |
| COXB | Cytochrome c oxidase polypeptide Vb | P10606 | 10 612 | 6,33 | 19 |
| COXE | Cytochrome c oxidase polypeptide VIa-liver | Q02221 | 9 619 | 6,42 | ND |
| COXG | Cytochrome c oxidasepolypeptide VIb | P14854 | 10 061 | 6.78 | 65 |
| CYC | Cytochrome c | P00001 | 11 617 | 9.59 | 62 |
| D3D2 | 3,2 trans enoyl CoA isomerase | P42126 | 28 736 | 6.00 | 23 |
| DHA X | Antiquitin | P49419 | 55 366 | 6.44 | 26 |
| DHE3 | Glutamate dehydrogenase | P00367 | 56 008 | 6.71 | 51 |
| DHSA | Succinate dehydrogenase flavoprotein subunit | P31040 | 68 012 | 6.25 | 21 |
| DLDH | Dihydrolipoamide dehydrogenase | P09622 | 50 148 | 6.35 | 11 |
| ECH1 | Delta3,5-delta2,4-dienoyl-CoA isomerase | Q13011 | 35 994 | 6.61 | 53 |
| ECHA | Trifunctional enzyme alpha subunit | P40939 | 78 970 | 8.98 | 34 |
| ECHB | Trifunctional enzyme beta subunit | P55084 | 47 485 | 9.24 | 42 |
| ECHM | Enoyl CoA hydratase, short chain | P30084 | 28 355 | 5.88 | 53 |
| EFTU | Elongation factor Tu | P49411 | 45 045 | 6.31 | 46 |
| ETFA | Electron transfer flavoprotein, alpha subunit | P13804 | 35 079 | 8.62 | 67 |
| ETFB | Electron transfer flavoprotein beta subunit | P38117 | 27 843 | 8.25 | 45 |
| ETFD | Electron transfer flavoprotein dehydrogenase | Q16134 | 64 676 | 6.52 | 32 |
| FUMH | Fumarate hydratase | P07954 | 50 082 | 6.99 | 32 |
| GPDM | Glycerol-3-phosphate dehydrogenase | P43304 | 76 323 | 6.17 | 23 |
| GR75 | Stress-70 protein | P38646 | 68 858 | 5.51 | ND |
| HCD2 | 3-hydroxyacyl-CoA dehydrogenase type II | Q99714 | 26 923 | 7.66 | 72 |
| IDHA | Isocitrate dehydrogenase [NAD] subunit alpha | P50213 | 36 640 | 5.71 | 24 |
| IVD | Isovaleryl-CoA dehydrogenase | P26440 | 43 069 | 6.90 | 20 |
| KAD3 | GTP:AMP phosphotransferase mitochondrial | Q9UIJ7 | 25 507 | 9.30 | 31 |
| KCRU | Creatine kinase, ubiquitous mitochondrial | P12532 | 43 080 | 7.31 | 44 |
| MDHM | Malate dehydrogenase | P40926 | 33 000 | 8.54 | 47 |
| MPPA | Mitochondrial processing peptidase alpha subunit | Q10713 | 55 625 | 5.75 | 32 |
| MTX2 | Metaxin 2 | O75431 | 29 763 | 5.90 | 32 |
| N7BM | NADH-ubiquinone oxidoreductase subunit B17.2 | Q9UI09 | 17 144 | 9.63 | 48 |
| NB7M | NADH-ubiquinone oxidoreductase B17 subunit | O95139 | 15 358 | 9.63 | 36 |
| NB8M | NADH-ubiquinone oxidoreductase B18 subunit | P17568 | 16 271 | 9.10 | 42 |
| NI2M | NADH-ubiquinone oxidoreductase B22 subunit | Q9Y6M9 | 21 700 | 8.59 | 71 |
| NI9M | NADH-ubiquinone oxidoreductase B9 subunit | O95167 | 9 279 | 7.98 | 41 |
| NIPM | NADH-ubiquinone oxidoreductase 15 kDa subunit | O43920 | 12 386 | 9.29 | 64 |
| NUAM | NADH-ubiquinone oxidoreductase 75 kDa subunit | P28331 | 77 053 | 5.36 | 42 |
| NUCM | NADH-ubiquinone oxidoreductase 49 kDa subunit | O75306 | 53869 | 7.21 | 41 |
| NUDM | NADH-ubiquinone oxidoreductase 42 kDa subunit | O95299 | 37 147 | 6.87 | 60 |
| NUEM | NADH-ubiquinone oxidoreductase 39 kDa subunit | AAH15837 (NCBI) | 38 899 | 9.70 | 41 |
| NUFM | NADH-ubiquinone oxidoreductase 13 kDa-B subunit | Q16718 | 13 327 | 5.76 | 44 |
| NUGM | NADH-ubiquinone oxidoreductase 30 kDa subunit | O75489 | 26 415 | 5.48 | 80 |
| NUHM | NADH-ubiquinone oxidoreductase 24 kDa subunit | P19404 | 23 760 | 5.71 | 53 |
| NUIM | NADH-ubiquinone oxidoreductase 23 kDa subunit | O00217 | 20 290 | 5.10 | ND |
| NUML | NADH-ubiquinone oxidoreductase MLRQ subunit | O00483 | 9 370 | 9.42 | 76 |
| NUPM | NADH-ubiquinone oxidoreductase 19 kDa subunit, PGIV | P51970 | 19 974 | 7.93 | 47 |
| OAT | ornithine amino transferase | P04181 | 44 808 | 5.72 | 21 |
| ODB2 | Lipoamide acyltransferase component of branched-chain alpha-keto acid dehydrogenase complex | P11182 | 46 710 | 6.40 | 45 |
| ODPA | Pyruvate dehydrogenase E1 component alpha subunit, somatic form | P08559 | 40 229 | 6.51 | 49 |
| ODPB | Pyruvate dehydrogenase E1 component beta subunit | P11177 | 35 890 | 5.38 | ND |
| ODO1 | 2-oxoglutarate dehedrogenase E1 component | Q02218 | 108 880 | 6.24 | 24 |
| ORN | Oligoribonuclease | Q9Y3B8 | 26 861 | 6.40 | 19 |
| PDX3 | Thioredoxin-dependent peroxide reductase | P30048 | 21 468 | 5.77 | 40 |
| POR1 | Voltage-dependent anion-selective channel protein 1 | P21796 | 30 641 | 8.63 | 58 |
| POR2 | Voltage-dependent anion selective channel protein 2 | P45880 | 38 639 | 6.32 | 41 |
| PPCM | Phosphoenolpyruvate carboxykinase | Q16822 | 67 005 | 6.58 | ND |
| PUT2 | Delta-1-pyrroline 5 carboxylate dehydrogenase | Q16882 | 59 066 | 6,96 | 42 |
| SODM | Superoxide dismutase [Mn] | P04179 | 22 204 | 6.86 | 75 |
| SSB | Single-strand DNA-binding protein | Q04837 | 15 195 | 8.24 | 43 |
| THTR | Thiosulfate sulfurtransferase, rhodanese | Q16762 | 33 298 | 6.83 | 53 |
| TRAL | Heat shock protein 75 kDa | Q12931 | 80 011 | 8.05 | |
| UCR1 | Ubiquinol-cytochrome C reductase complex core protein I | P31930 | 49 102 | 5.43 | 46 |
| UCR6 | Ubiqinol-cytochrome C reductase complex 14 kDa protein | P14927 | 13 399 | 8.73 | 43 |
| UCRI | Ubiqinol-cytochrome C reductase iron-sulfur subunit, Rieske protein | P47985 | 21 617 | 6.30 | 35 |
| | Isocitrate dehydrogenase 2 (NADP+) | Q96GT3 | 50 909 | 8.88 | 45 |
| | Mitochondrial thioredoxin reductase | Q9H2Z5 | 56 204 | 7.24 | ND |
| | Mitofilin | Q9P0V2 | 68 187 | 5.57 | ND |
| | NADH-cytochrome b5 reductase isoform | Q9UHQ9 | 34 095 | 9.41 | 45 |
| | NADH-ubiquinone oxidoreductase B16.6 subunit | Q9P0J0 | 16 698 | 8.04 | 41 |
| | novel AMP-binding enzyme similar to acetyl-coenzyme A synthethase (acetate-coA ligase) | Q9NUB1 | 52 592 | 5.85 | ND |
| | Similar to tricarboxylate carrier-like protein | Q9BWM7 | 35 823 | 9.26 | 50 |
| | Ubiquinol-cytochrome c reductase core protein II | Q9BQ05 | 48 443 | 8.74 | 40 |

| Code | Name | Accession number | mass | pI | Sequence coverage (%) |
|---|---|---|---|---|---|
| | CONTAMINANTS | | | | |
| 143Z | 14-3-3 zeta/delta | P29312 | 27 745 | 4.73 | 24 |
| ACTB | Actin, cytoplasmic 1 | P02570 | 41 606 | 5.29 | ND |
| ACTZ | Alpha-centractin | P42024 | 42 613 | 6.19 | 50 |
| ALBU | Serum albumin | P02768 | 66 472 | 5.67 | 44 |
| ANX1 | Anexin I (fragment) | P04083 | 38 787 | 6.64 | 16 |
| ANX2 | Anexin II | P07355 | 38 472 | 7.56 | 52 |
| AOP2 | Antioxidant protein 2 | P30041 | 24 904 | 6.02 | ND |
| AOPP | Putative peroxisomal antioxidant enzyme | P30044 | 16 863 | 6.84 | 30 |
| AR20 | ARP2/3 complex 20 kDa subunit | O15509 | 19 667 | 8.53 | 36 |
| ARP3 | Actin-like protein 3 | P32391 | 47 371 | 5.61 | 55 |
| COF1 | Cofilin, non muscle isoform | P23528 | 18502 | 8.22 | 38 |
| CRAB | Alpha crystallin B chain | P02511 | 20 159 | 6.76 | 40 |
| CYPB | Peptidyl-propyl cis-trans isomerase B | P23284 | 20 289 | 9.25 | 33 |
| ENPL | Endoplasmin | P14625 | 90 178 | 4.73 | 51 |
| EZRI | Ezrin (fragment) | P15311 | 69 267 | 5.95 | 18 |
| FIBB | Fibrinogen beta chain | P02675 | 55 928 | 8.54 | 67 |
| FLT1 | Flotillin-1 | O75955 | 47 332 | 7.04 | 49 |
| GR78 | 78 kDa glucose-regulated protein | P11021 | 70 479 | 5.01 | 51 |
| GTP | glutathione S transferase P | P09211 | 23 224 | 5.44 | 56 |
| H4 | Histone H4 (fragment) | P02304 | 11 236 | 11.36 | 30 |
| K2C1 | Keratin type II cytoskeletal 1 (fragment) | P04264 | 65 886 | 8.16 | 17 |
| PDA3 | Protein disulfide isomerase A3, ERP60 | P30101 | 57 146 | 5.98 | 51 |
| PDA6 | protein disulfide isomerase A6 | Q15084 | 46 171 | 4.95 | 17 |
| PDI | Protein disulfide isomerase | P07237 | 55 294 | 4.69 | 43 |
| PDX4 | Peroxiredoxin 4 | Q13162 | 30 540 | 5.86 | 49 |
| PERM | Myeloperoxidase | P05164 | 66 107 | 9.22 | 19 |
| PHB | Prohibitin | P35232 | 29 804 | 5.57 | 80 |
| PLL | Lactogen, Choriomammotrophin | P01243 | 22 308 | 5.33 | ND |
| PSA7 | Proteasome subunit alpha type 7 | O14818 | 28 041 | 8.60 | 17 |
| S110 | Calpactin I light chain | P08206 | 11 179 | 7.3 | 52 |
| TRFE | Serotransferrin | P02787 | 75 181 | 6.7 | 43 |
| | B-Cell receptor associated protein, D-prohibitin | Q99623 | 33 296 | 9.83 | 59 |
| | Hemoglobin alpha 1 globin chain (fragment) | Q9BX83 | 10703 | 7.07 | 40 |
| | Human immunoglobin heavy chain variable region (fragment) | U00530 (OWL) | 12 099 | 8.64 | 40 |
| | PI 5.3 beta 2-microglobulin | Q9UDF4 | 11 618 | 6.07 | 55 |
| | SUBCELLULAR LOCALIZATION UNKNOWN / POORLY CHARACTERIZED PROTEINS | | | | |
| | AD025 | Q9HBL7 | 17 190 | 9.58 | 34 |
| BAF | Barrier-to-autointegration factor | O75 531 | 10 059 | 5.81 | 65 |
| | Brain my025 | Q9H3J9 | 20 199 | 9.35 | 36 |
| CG51 | Protein CGI-51 | Q9Y512 | 51 976 | 6.44 | 43 |
| | Homeotic protein Hox-4.8 (fragment) | A56563 | 9 889 | 10.41 | 45 |
| | Hypothetical 14.5 kDa protein | O95329 | 14 635 | 5.86 | 25 |
| | Hypothetical 23.7 kDa protein (Fragment) | Q9BVM2 | 23 847 | 9.1 | 23 |
| | Putative transposase A | AAK50868 | 10 182 | 9.79 | 43 |
| | SH3BGRL2-like protein | Q9BPY5 | 12 326 | 6.29 | 70 |
| | Similar to Acetyl-CoA carboxylase beta subunit | BAB74063 (NCBI) | 35 718 | 7.67 | 14 |
| | Uncharacterized hematopoietic stem/progenitor cells protein MDS030 | Q9NZ44 | 17 218 | 9.69 | 34 |
| | Uncharacterized hypothalamus protein HT012, PNAS-27 | Q9NPJ3 | 14 960 | 9.23 | 50 |

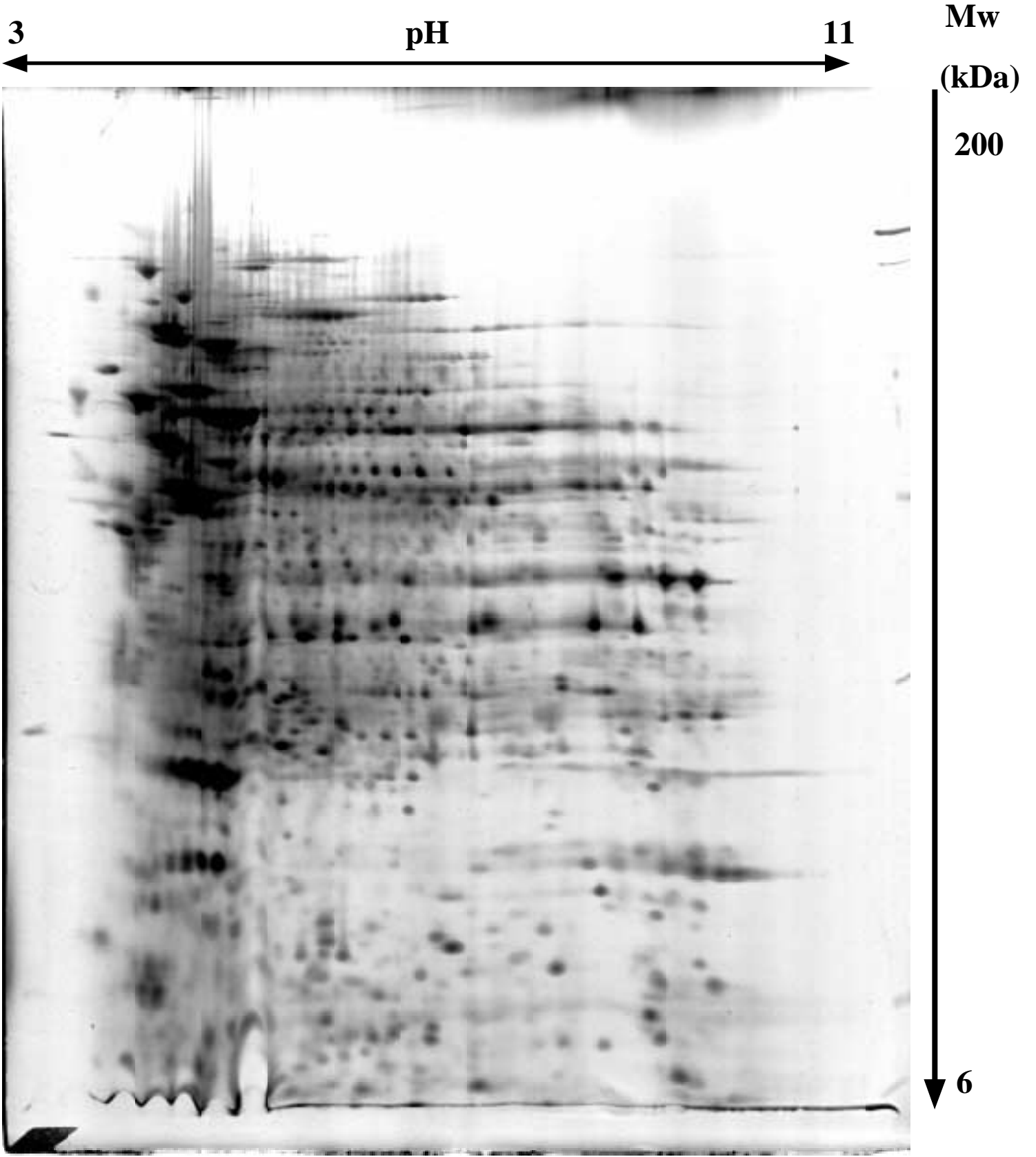

Figure 2

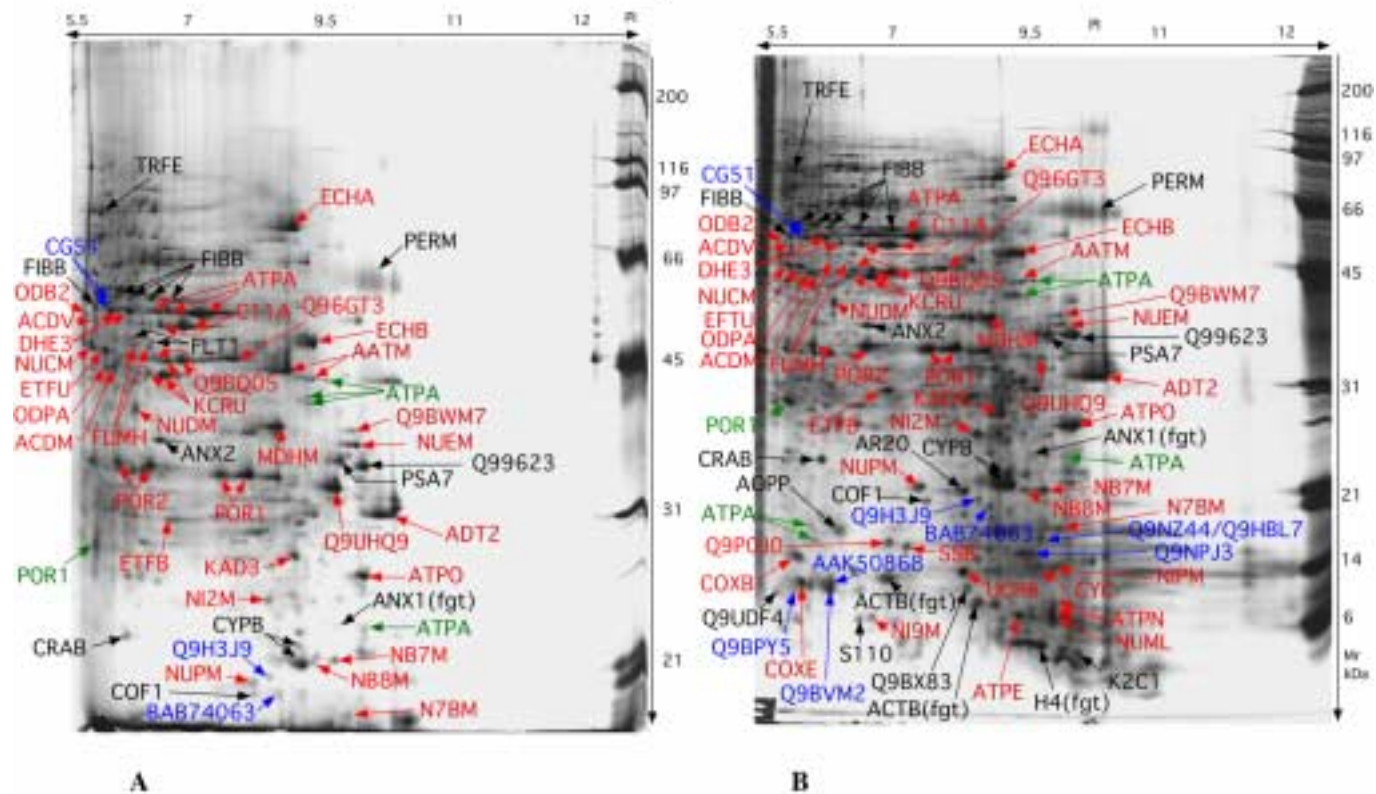

Figure 3

**Figure 4**

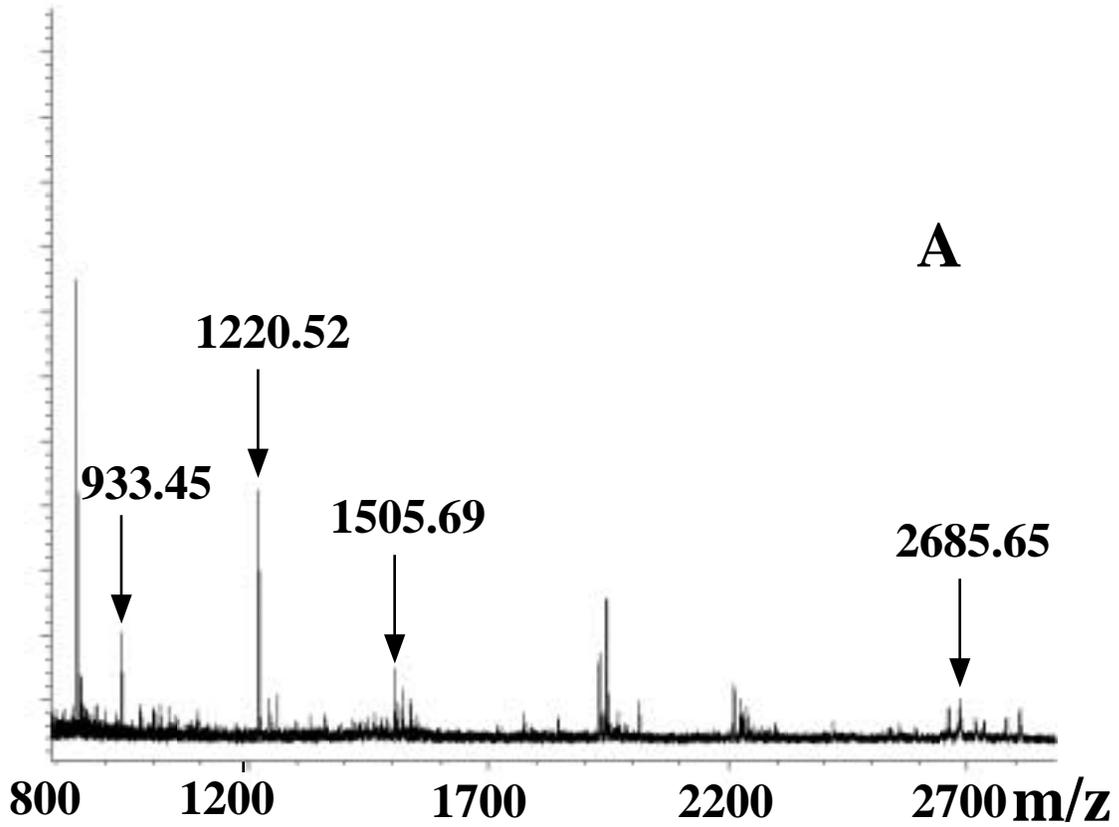

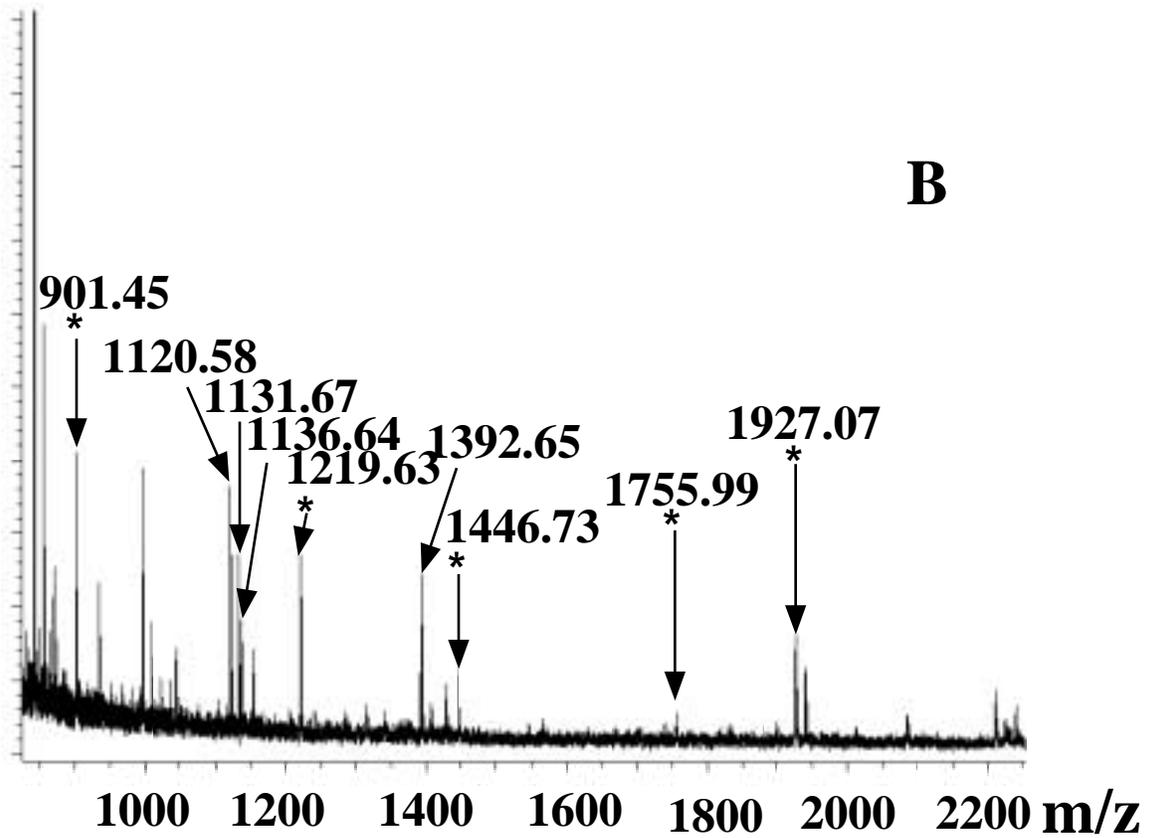